\documentclass[aps,prb,twocolumn,10pt, floatfix]{revtex4-1}
\usepackage{graphicx}
\usepackage{amssymb}
\usepackage{amsmath}
\newcommand\numberthis{\addtocounter{equation}{1}\tag{\theequation}}
\usepackage{color}
\usepackage[dvipsnames]{xcolor}
\usepackage{slantsc}
\usepackage{float}
\graphicspath{{./FIG_FINALI/}}
\usepackage{bbold}
\usepackage[caption=false]{subfig}
\usepackage{array}
\usepackage{braket}
\usepackage{hyperref}
\hypersetup{backref,pdfpagemode=FullScreen,colorlinks=true,allcolors=blue}

\newcommand{\rededit}[1]{\textcolor{black}{#1}}

\begin{document}
\title{Wannier-function-based constrained DFT with nonorthogonality-correcting Pulay forces in application  
to the reorganization effects in graphene-adsorbed pentacene}
\author{Subhayan Roychoudhury}
\affiliation{School of Physics, AMBER and CRANN Institute, Trinity College Dublin, Dublin 2, Ireland}
\author{David D. O'Regan}
\email{david.o.regan@tcd.ie}
\affiliation{School of Physics, AMBER and CRANN Institute, Trinity College Dublin, Dublin 2, Ireland}
\author{Stefano Sanvito}
\affiliation{School of Physics, AMBER and CRANN Institute, Trinity College Dublin, Dublin 2, Ireland}

\begin{abstract}
Pulay terms arise in the Hellman-Feynman forces in electronic structure calculations when one employs a basis set made of 
localized orbitals that move with their host atoms. If the total energy of the system depends on a subspace population defined 
in terms of the localized orbitals across multiple atoms, then unconventional Pulay terms will emerge due to the variation of the 
orbital nonorthogonality with ionic translation. Here, we derive the required exact expressions for such terms, which cannot be
eliminated by orbital orthonormalization.
We have implemented these corrected ionic forces within the linear-scaling density functional theory (DFT) package {\sc onetep}, 
and have used constrained DFT to calculate the reorganization energy of a pentacene molecule adsorbed on a graphene flake. The 
calculations are performed by including ensemble DFT, corrections for periodic boundary conditions, and empirical Van der Waals 
interactions. For this system we find that tensorially invariant population analysis yields an adsorbate subspace population that is 
very close to integer-valued when based upon nonorthogonal Wannier functions, and also but less precisely when using pseudoatomic 
functions. Thus, orbitals can provide a very effective population analysis for constrained DFT. Our calculations show that the reorganization 
energy of the adsorbed pentacene is typically lower than that of pentacene in the gas phase. We attribute this effect to steric hindrance.
\end{abstract}

\maketitle

\section{Introduction}

Across a wide range of electronic structure theory methods, such as constrained density functional theory (cDFT)~\cite{WVV,KadukB_KT_VT}, 
density functional theory plus Hubbard $U$ (DFT+$U$)~\cite{PhysRevB.44.943,QUA:QUA24521}, DFT combined with dynamical 
mean field theory (DMFT)~\cite{0953-8984-9-35-010,RevModPhys.78.865}, wave function-embedding~\cite{SimonB_MS_TM_FManby,SherwoodP_HV_A_JCS_PGS_ABN_AVM_HHI}, 
and some perturbative approaches in quantum chemistry~\cite{JCC:JCC23150}, 
the population of a particular subspace is physically relevant and the total energy depends explicitly upon it. Thus, the ability to define 
appropriate subspaces for population analysis is of considerable importance. This is exemplified by the sustained efforts in recent years 
in the development of physically-motivated orbitals such as maximally localized Wannier functions (MLWF)~\cite{PhysRevB.56.12847}, 
nonorthogonal localized molecular orbitals (NOLMO)~\cite{C3CP52989D}, muffin-tin orbitals (MTO)~\cite{2000LNP...535....3A}, and 
natural bond orbitals (NBO)~\cite{B1RP90011K} for use in system-dependent, adaptive population analysis. Population analysis by 
means of projection onto orbitally-defined subspaces has undergone detailed analysis in recent years~\cite{PhysRevB.90.075128,0953-8984-27-24-245606} and, in particular, the effects of projector orbital ambiguity in DFT+$U$~\cite{PhysRevB.82.081102,YueChaoWang_ZHC_HJ} and DFT+DMFT~\cite{PhysRevB.81.195107} have been investigated in detail.

In calculations in which the total energy depends explicitly upon localized  orbitals that are centred on atoms, Pulay 
terms~\cite{doi:10.1021/ja00504a009,PhysRevB.48.4937} arise in the Hellmann-Feynman forces due to spatial translations of 
the orbitals. It is, however, less known, although previously identified~\cite{WuVoorhisForce,DNDmMKVIA}, that additional Pulay 
terms emerge when the total energy also depends on the overlap matrix of such orbitals. This is  necessary
 for correct 
population analysis using nonorthogonal orbitals. In fact, these forces are present for any multi-centre atomic projection of the 
density or the Kohn-Sham density matrix. They exist  when using orthonormal orbitals such as MLWFs, for example, since any ionic 
movement typically breaks the orthornormality. Thus, unless the forces take into account that the orbitals are regenerated or 
orthonormalized following a translation, a condition which is difficult to encode, then unconventional \emph{nonorthogonality 
Pulay forces} arise even for  orbitals that are defined as orthonormal.

Approaches for calculating the necessary corrections, based on a L\"{o}wdin orthonormalized representation of the subspace 
projection, invariably encounter a cumbersome, difficult to solve, Sylvester equation~\cite{Bartels:1972:SME:361573.361582} of the form 
\begin{align} 
\frac{ d \mathbf{O} 
}{ d R_i }
&{}=
\mathbf{O}^{1/2} \frac{ d \mathbf{O}^{1/2} 
}{ d R_i } + 
\frac{ d \mathbf{O}^{1/2} 
}{ d R_i }  \mathbf{O}^{1/2} \nonumber \quad
\mbox{or} \\
-   \frac{ d \mathbf{O} 
}{ d R_i } 
&{}=
\mathbf{O}^{1/2} \frac{ d \mathbf{O}^{-1/2} 
}{ d R_i } + 
\frac{ d \mathbf{O}^{-1/2} 
}{ d R_i }  \mathbf{O}^{1/2}   , \end{align} 
where $\mathbf{O}$ is the projector orbital overlap matrix and $R_i$ is a Cartesian component of the ionic position. Here
the solution for $\mathbf{O}^{1/2}$ is required. An approximate method for working around this problem, based on neglecting 
off-diagonal matrix elements in $\mathbf{O}^{-1/2}$, has been recently proposed in reference~[\onlinecite{DNDmMKVIA}]. 
Reference~[\onlinecite{WuVoorhisForce}] instead provides a formula for the full matrix $d \mathbf{O}^{1/2} / d R_i$, which 
makes use of the basis of the shared eigenvectors of $\mathbf{O}$ and $\mathbf{O}^{1/2}$. This necessitates matrix 
diagonalization. The applicability and practicality of these two approaches depend on the  details of the force 
calculations to be undertaken. 

\rededit{In this work we use nonorthogonal basis functions and their appropriate tensor notation following a long standing tradition in electronic structure theory~\cite{doi:10.1063/1.1740588, doi:10.1063/1.437441, doi:10.1063/1.475423, PhysRevA.43.5770, PhysRevB.66.035119, PhysRevB.95.115155}. We furthermore use the modern tensorially invariant population analysis~\cite{ORPM}, which has appeared in various contexts~\cite{PhysRevB.90.075128, 0953-8984-27-24-245606} including that of cDFT~\cite{PhysRevB.93.165102, Lukman}. We extend this to calculate an exact, simple and intuitive expression for the nonorthogonality Pulay  forces, 
which circumvents orbital orthonormalization and overlap matrix diagonalization entirely.}  This expression is applicable to real and complex valued orbitals alike, and whether or not they are orthonormal at the point of 
force evaluation. Avoiding matrix diagonalization ensures its applicability to large systems using 
linear-scaling DFT. Our scheme is put here to the test by calculating the reorganization energy of a pentacene molecule 
physisorbed on a graphene sheet.

The paper is organized as follows. In the next section we will define the physical problem addressed by our work, namely the 
calculation of the energies needed for extracting the reorganization energy of a molecular absorbate on a metallic substrate.
Then we will move on describing our computational methods,  focussing on the derivation of the forces in orbital-based
cDFT, the performance of orbital-based population analysis, and a number of practical considerations addressed using the 
{\sc onetep} code. Our results for pentacene on graphene will be presented next, followed by our conclusions.

\section{Physical problem: reorganization of a charged molecule physisorbed on a metallic surface}
The reorganization energy holds paramount importance in charge transport calculations. Semi-classical Marcus theory~\cite{JofEC} 
at high temperature, $T$, computes the probability per unit time of an electron hopping, $k_\textrm{ET}$, from the Fermi's 
Golden Rule as~\cite{doi:10.1021/jp9605663,doi:10.1021/ct200388s}
\begin{equation}\label{FGR}
\begin{split}
k_\textrm{ET}=\frac{\lvert \bra{i}\hat{H}\ket{f} \rvert^2}{\hbar}\sqrt{\frac{\pi}{\lambda k_B T}}\exp\left[-\frac{(\lambda+\Delta G^0)^2}{4\lambda k_\textrm{B} T}\right]\:,
\end{split}
\end{equation}
where $\hat{H}$ is the Hamiltonian, $\ket{i}$ and $\ket{f}$ are the initial and final electronic states, respectively, and $\Delta G^0$ 
is the change in Gibbs' free energy associated to the charge transfer process. The reorganization energy, which enters the exponential
term defining $k_\textrm{ET}$, is thus an important ingredient~\cite{C4CP06078D,doi:10.1021/jp9535250, doi:10.1021/jp4046935} for the calculation of the charge hopping. In this work we compute the 
reorganization energy of a pentacene molecule. In its crystalline solid state form, due to its high HOMO level, pentacene is a $p$-type
semiconductor~\cite{YoshiroYamashita} with a high hole mobility~\cite{THJT}. Thus, ionization reorganization effects in pentacene-based 
systems are of significant interest, being the subject of several theoretical and experimental 
studies~\cite{Gruhn_SF_B_T_M_M_C_K_B,CMFGBB,SC_CFDFOMSSB}. 

\begin{figure}
\centering
\includegraphics[width=\columnwidth]{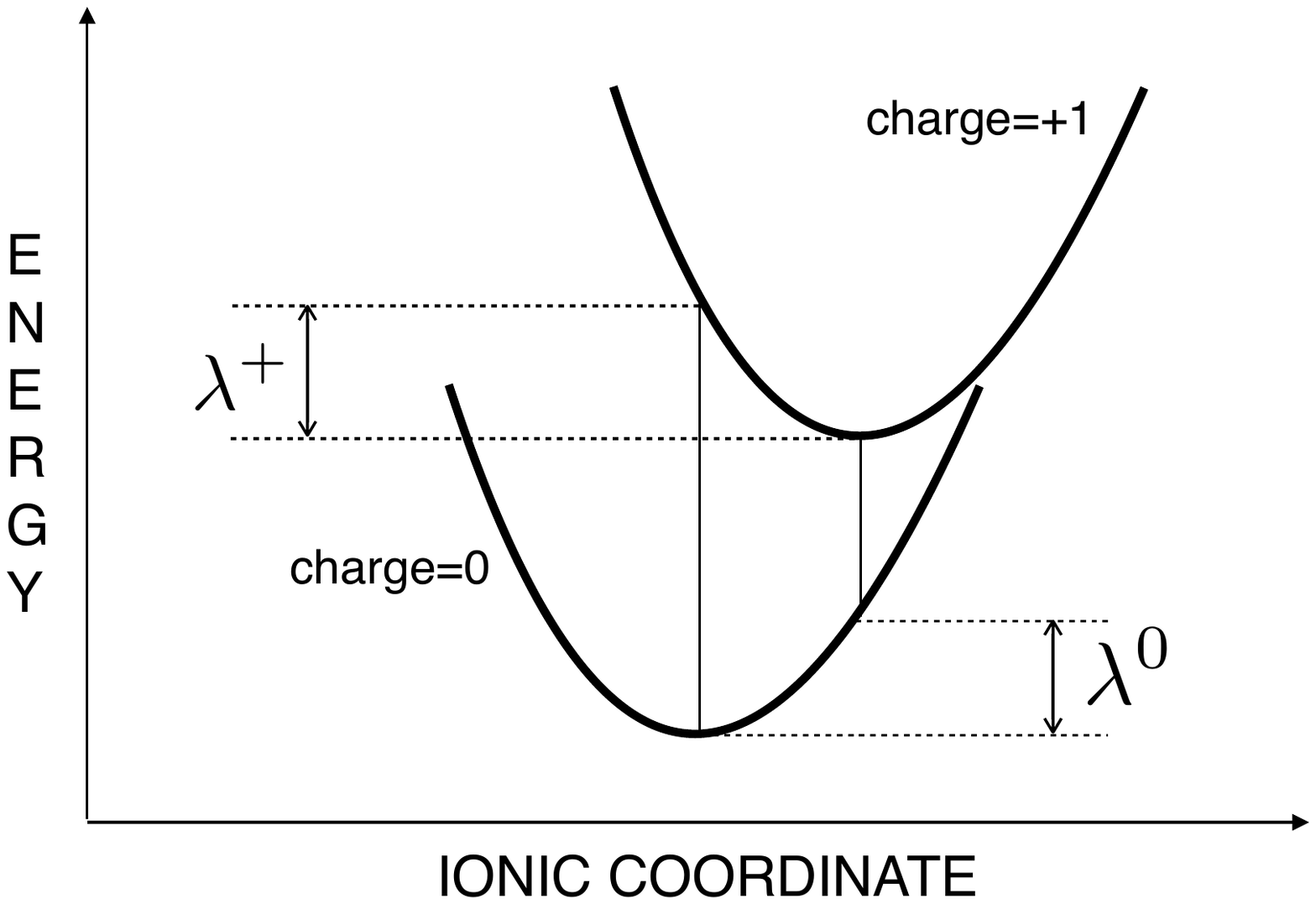}
\caption {Schematic diagram of the energy as a function of the ionic coordinates for a charged and a neutral molecule. The 
reorganization energy is defined as $\lambda^0+\lambda^+$. }\label{FIG:ReorgEnDiagram}
\end{figure}

Let us  define the reorganization energy precisely.
The ionic coordinates of any system depend on its electronic occupation. For instance, if an electron is removed from a 
neutral molecule, such as in photoemission spectroscopy, its ionic coordinates will readjust to a new geometry due to the 
local electron-phonon coupling~\cite{doi:10.1021/jp900157p,doi:10.1021/cr050140x}. Figure~\ref{FIG:ReorgEnDiagram} 
shows two parabolic curves corresponding to the energy surface of the neutral molecule and that of the singly ionized
one as a function of some collective atomic coordinates. We define as $\lambda^0$ the energy difference between 
the ground state geometry and the ground state geometry of the charged configuration~\cite{C0CS00198H}, when the
molecule is neutral. In contrast, $\lambda^+$ is the same quantity but calculated for the ionized system. The reorganization 
energy, $\lambda$, for the molecule undergoing electron removal is defined as
\begin{equation}\label{equReOr}
\lambda=\lambda^0+\lambda^+\:,
\end{equation}
where similar definitions can be given for the case where the molecule receives an extra electron.

Theoretical approaches to compute the reorganization energy typically consist of either calculating the energy difference from the 
adiabatic potential energy surface, or of indirectly evaluating the molecule's normal modes ~\cite{YX_WL_WC_LW_SZ}. Here we 
adopt the former approach. For an isolated pentacene molecule, an electron removal can be simulated with unconstrained DFT and 
therefore it does not require the aforementioned force terms. However, this approach is not viable for the study of reorganisation in 
systems relevant to organic semiconductor devices, where organic molecules is typically adsorbed on metallic electrodes. When a 
molecule is adsorbed on a metallic substrate and its highest occupied molecular orbital (HOMO) lies below the Fermi level, the hole
must be prevented from migrating to the energetically favourable location of the substrate. We achieve this by using cDFT to force 
the hole onto the adsorbate.

cDFT has been widely applied to the study of charge transfer in organic compounds~\cite{Franke2012,doi:10.1021/acs.jctc.6b00815,doi:10.1063/1.4979895,doi:10.1021/acs.jctc.5b00057,C6SC04547B,PhysRevB.93.195208,doi:10.1021/acs.jctc.6b00564,doi:10.1021/acs.jctc.6b01085,doi:10.1080/00268976.2013.800243}.
Recently, cDFT has been used to estimate charge-transfer excitations in bulk pentance in the infinite-crystal limit~\cite{PhysRevB.93.165102}. 
The present work utilises the same underlying linear-scaling cDFT implementation, itself an extension of a linear-scaling implementation of 
DFT+$U$~\cite{PhysRevB.85.085107} using nonorthogonal generalized Wannier functions. Also relevant to this work is that cDFT 
has been used to simulate removal or addition of electrons from adsorbed molecule in the context of calculating charge transfer 
energies~\cite{PhysRevB.88.165112,PhysRevB.93.045130}. Here we use cDFT in conjunction with nonorthogonality Pulay forces 
to calculate the reorganization energy of a pentacene molecule physisorbed on a flake of graphene. \rededit{The energy of a system as a function of its geometry can contain multiple local minima, and this is particularly  the case
for the incommensurate corrugated system at hand. 
The proposed method, in conjunction with efficient sampling techniques like simulated annealing~\cite{Kirkpatrick671}, basin hopping~\cite{doi:10.1021/jp970984n}, etc., could be used to explore such 
energy landscapes in the presence
of orbital-based constraints.} We note that a more complex system, 
consisting of a film of weakly bound pentacene molecules adsorbed on highly oriented pyrolytic graphite (HOPG), has been the 
subject of several theoretical and experimental studies \cite{KERA2009135,YH_NS_FH_KS_FR_OKK_UN,PP_CV_BJL}. It has 
been shown, in the experimental work of reference~[\onlinecite{KERA2009135}], that the reorganization energy of 
pentacene is there, remarkably, higher than that in the gas phase.

\section{Theoretical Problem: population analysis and forces based on nonorthogonal orbitals}

In cDFT, to date, real-space partitioning has prevailed over orbital-based population analysis methods. 
Central to the viability of using more chemically motivated orbitals to define the constrained
population in cDFT, and perhaps hindering their adoption, is the proper treatment of their nonorthogonality.
In particular, historically there has been some uncertainty~\cite{2008JPCM...20F5205T,PhysRevB.73.045110} 
as to how subspace populations should be defined in terms of nonorthogonal orbitals, which typically
(but not necessarily by any means) form a subset of the basis set for the Kohn-Sham states.
This uncertainty has previously been conclusively resolved within the context of DFT+$U$~\cite{ORPM}, and the correct 
procedure has recently been pioneered in cDFT for calculating charge-transfer energies in solid pentacene~\cite{PhysRevB.93.165102}.
We will numerically investigate the performance of this tensorially consistent procedure  for cDFT in the present work.
A separate problem, which we also will touch upon in this work, is the arbitrary choice of the underlying projection orbitals  
in terms of their particular spatial profile.

The canonical orbital-based population analyses in quantum chemistry are due to L\"{o}wdin and Mulliken, and these are both 
unsuitable for cDFT.
cDFT population analyses based on orbitals that are globally L\"{o}wdin orthornormalized, meaning that the entire
basis set is orthonormalized before a subset is selected out, typically collect density contributions from all atoms in 
the simulation cell, regardless of how distant they may be from the region of interest. Mulliken  population analysis, 
constructed using the global orbital overlap matrix has the same problem~\cite{ORPM}. In a nutshell both methods 
have the fundamental difficulty that the measured population is arbitrary with respect to linear transformations among 
the selected subset of the orbitals (an example of broken \emph{tensorial invariance}).

A tensorially invariant population analysis~\cite{PhysRevB.93.165102,ORPM} instead gathers density contributions and 
applies constraining potentials only within the region of interest. We will here demonstrate that this can provide very 
reasonable electronic populations for a physisorbed molecule when using either pseudoatomic orbitals or generalized 
Wannier functions. Physically-motivated orbitals can thus compete with real-space weight functions in cDFT, when treated 
appropriately. Their use may be particularly advantageous in situations where the system or observable of interest does not 
readily admit a real-space partitioning, such as when constraining the population of an atom in a crystal, or that of a group 
of single-particle states based on their principal angular momentum character. 

\section{Methodology}

\subsection{Constrained density-functional theory forces}

In DFT the ground state (GS) electron density, $\rho_0$, uniquely specifies all the GS properties of a system, including 
its GS energy~\cite{HPKW}. This can thus be found by variationally minimizing an approximate energy functional, $E\left[\hat{\rho}\right]$, 
where $\hat{\rho}$ is the density operator. In cDFT, instead, one seeks to find the GS of the system subject to a constraint, 
for example the constraint that a given number of electrons is found in a particular subspace. This simple constraint 
has the mathematical form 
\begin{equation}\label{equ2_ReorgEn}
{\rm Tr} [\hat{\rho}\hat{\mathbb{P}} ] - N_\textit{c} = 0, 
\end{equation}
where $\hat{\mathbb{P}}$ is projection operator for the subspace of interest and $N_c$ is the target number of electrons (here
`Tr' indicates the trace of the operator, computed over an appropriate basis set). In order to find the density corresponding to such 
constrained ground state, one finds the stationary point~\cite{WVV,ORT} of the functional $W\left[\hat{\rho},V_\textit{c}\right]$, 
where $V_\textit{c}$ is a Lagrange multiplier and
\begin{equation}\label{equ3}
W\left[\hat{\rho},V_\textit{c}\right]=E\left[\hat{\rho}\right]+V_\textit{c}\left({\rm Tr}[\hat{\rho}\hat{\mathbb{P}}]-N_\textit{c}\right).
\end{equation}
For a given $V_\textit{c}$ the Kohn-Sham potential is modified by the addition of the term $V_\textit{c} \hat{\mathbb{P}}$ and 
$W \left[\hat{\rho},V_\textit{c} \right]$ is minimized as a functional of $\hat{\rho}$ as usual. Considering just the global 
minima for each $V_\textit{c}$, $W$ can be regarded as a function $W\left(V_\textit{c}\right)$ of $V_\textit{c}$ alone~\cite{WVV} 
(strictly speaking, constrained systems can be constructed where it is a multiple-valued non-function~\cite{ORT}).
The stationary points of $W\left(V_\textit{c}\right)$ yield the (potentially degenerate) ground state densities of the system subject to the 
given constraint. In particular, the stability of a ground-state ensures that $W\left(V_\textit{c}\right)$ attains a maximum~\cite{ORT} 
with respect to $V_\textit{c}$. At the stationary point $W\left[\hat{\rho},V_\textit{c}\right]=E\left[\hat{\rho}\right]$, 
since Eq.~(\ref{equ3}) is satisfied. 
 
In general, $E\left[\hat{\rho}\right]$ is not stationary at a non-trivially constrained density and, hence, the Hellmann-Feynman 
theorem cannot be applied to $E\left[\hat{\rho}\right]$ alone. It is instead applied to $W\left[\hat{\rho},V_\textit{c}\right]$ 
in order to find the ionic force 
\begin{equation}\label{equ4}
F_i=-\frac{dW}{d R_i}=
-\frac{\partial W}{\partial R_i}
- {\rm Tr} \left[ \frac{\partial W}{\partial \hat{\rho}} \frac{{\rm d} \hat{\rho}}{{\rm d} R_i} \right]
- \frac{\partial W}{\partial V_\textit{c} } \frac{{\rm d} V_\textit{c} }{{\rm d} R_i} , 
\end{equation}
where the index `$i$' is collective for the ion number and the Cartesian direction indexes. Here the term containing the trace 
vanishes at any stable ground-state by virtue of the Hellmann-Feynman theorem~\cite{ORT}, and the final term vanishes at 
the cDFT stationary points, i.e., where $\partial W / \partial V_\textit{c} = 0$.
The force is thus given, in practice,  by
\begin{equation}\label{equ5_ReOr}
F_i=
-\frac{\partial W}{\partial R_i} = 
-\frac{\partial E\left[\hat{\rho}\right]}{\partial R_i}-V_\textit{c}{\rm Tr} \left[\hat{\rho} \frac{\partial \hat{\mathbb{P}}}{\partial R_i} \right]\:.
\end{equation}
The first term on the right hand side is the contribution from the conventional DFT external potential of the constrained 
density~\cite{doi:10.1063/1.4728026}, 
while the second term, which we will denote by $F_i^\textit{c}$, is the  Pulay force due to the constraint. 
Before evaluating this contribution, we must next discuss how the subspace projection operator $\hat{\mathbb{P}}$ is constructed.
 
\subsection{Tensorially invariant population analysis}

When defining $\hat{\mathbb{P}}$ in terms of nonorthogonal orbitals, such as atomic orbitals centred on atoms, let us label them 
$\lvert \phi_m \rangle$, it is a commonplace and usually unnecessary practice to orthonormalize them by L\"{o}wdin transformation.
This generates orbitals of the form $ \lvert \tilde{\phi_j} \rangle = \sum_m \lvert \phi_m \rangle O^{-\frac{1}{2}}_{m j} $, 
where  $\mathbf{O}$ is an orbital overlap matrix. The matrix fractional power is most easily calculated by diagonalizing $\mathbf{O}$, 
taking the corresponding power of the eigenvalues, and by performing the inverse of the original diagonalizing transformation to 
arrive at $\mathbf{O}^{-\frac{1}{2}}$.

In methods dealing with the population of orbital-based subspaces such as cDFT and DFT+$U$,  it has been shown~\cite{ORPM} to 
be quite incorrect to use for $\mathbf{O}$ the overlap matrix $\mathbf{S}$ of any larger set that the projector orbitals $ \lvert \phi_m \rangle $ 
may be chosen from, since then the orthonormalized functions $ \lvert \tilde{\phi_j} \rangle $ extend across the larger subspace.
Instead, if the projection orbitals $ \lvert \phi_m \rangle $ used to span a cDFT subspace happen to be selected from a larger set of basis 
orbitals (e.g. the one spanning the entire Kohn-Sham space), then the subspace overlap matrix $\mathbf{O}$ must be extracted as a sub-block 
from the full overlap matrix $\mathbf{S}$ \emph{before} being diagonalized~\cite{PhysRevB.93.165102}.

As an example, let us imagine a bipartite system composed of natural but non-trivially overlapping source and drain regions 
for a charge-transfer excitation to be accessed using cDFT. If the source-region orbitals $ \lvert \tilde{\phi_j} \rangle $ are built 
using $\mathbf{S}$, then they will extend to some amount over the drain region, and vice-versa, in an uncontrolled manner.
This pathology will not arise if separate, smaller subspace overlap matrices $\mathbf{O}$ are defined for each of the source and 
drain regions. This also ensures tensorial invariance and, in particular, physical occupancy eigenvalues  (i.e., $0 \le \lambda_j \le 1$) 
for the projected density matrices of each constrained subspace~\cite{ORPM}.

By defining the subspace population as the trace over such orthonormalized functions, we obtain
\begin{align*}
\rm {Tr} \big[\hat{\rho}\hat{\mathbb{P}} \big]
&{}=\sum_{j }\braket{\tilde{\phi}_j |\hat{\rho}|\tilde{\phi}_j}\\
&{}=\sum_{m n}\braket{\phi_m|\hat{\rho}|\phi_n}\sum_{j} O^{-\frac{1}{2}}_{n j } O^{-\frac{1}{2}}_{j m}\:. \numberthis \label{equLOWD2}
\end{align*}
Equation (\ref{equLOWD2}) suggests a straightforward alternative approach, albeit one that is not available if the index
$j$ does not run over the same orbital count as $n$ and $m$ (such as when the delocalizing global matrix $\mathbf{S}$ is used).
Instead of performing a L\"{o}wdin orthonormalization, we may accept the nonorthogonality of the projectors and define the 
subspace population as a tensor contraction over the nonorthogonal set of $ \ket{\varphi_m} $ and their biorthogonal \rededit{complements}
$\ket{\varphi^m}$, defined through $\braket{\varphi^m|\varphi_{n}}=\delta^m_{\; n}$. This gives the transformations
\begin{equation}\label{equ6}
 \ket{\varphi^m}=\ket{\varphi_{n}}O^{nm} \quad
 \Leftrightarrow \quad
 \ket{\varphi_m}=\ket{\varphi^{n}}O_{nm}\:,
\end{equation}
where we have adopted the Einstein summation convention for contracting over paired indices, and where $O^{mn}$ is an element 
of the matrix $\mathbf{O}^{-1}$ and $O_{mn}=\langle \varphi_m \rvert \varphi_n\rangle$. If the functions $ \ket{\varphi_m} $ are chosen 
to be localized over a particular spatial region then also the functions $\ket{\varphi^m}$ will be. The required subspace occupancy is 
then given by
\begin{align*}
\textrm{Tr} \big[
\hat{\rho}\hat{\mathbb{P}}\big] &{} =
\textrm{Tr} \big[
\hat{\rho}\ket{\varphi^m}  \bra{\varphi_m}
 \big] = 
\bra{\varphi_m}\hat{\rho}\ket{\varphi_{n}}O^{nm}\numberthis\:, \label{equ7_RO}
\end{align*}
which is equivalent to Eq.~(\ref{equLOWD2}). Next, we look at how the  Pulay force  of cDFT appears when we make this 
simplification, i.e., when we use the contraction $\sum_j O^{-\frac{1}{2}}_{n j} O^{-\frac{1}{2}}_{j m}=O^{n m}$ \emph{before} 
the ionic-position derivative is taken.

\subsection{The nonorthogonality Pulay forces}

A change in the degree of nonorthogonality between projecting orbitals centred on atoms is a natural occurrence in calculations 
involving ionic displacements. In order to account for this, the final term of Eq.~(\ref{equ5_ReOr}) may be expanded, in view 
of Eq.~(\ref{equ7_RO}), as
\begin{align*}
 F_i^\textit{c}  =&{} -V_\textit{c} \Bigg[ \braket{\frac{\partial\varphi_m}{\partial R_i}|\hat{\rho}|\varphi_{n}}O^{nm}+\braket{\varphi_{m}|\hat{\rho}|\frac{\partial\varphi_{n}}{\partial R_i}}O^{nm}+ \\
&{} \quad + \braket{\varphi_m|\hat{\rho}|\varphi_{n}}\frac{\partial O^{nm}}{\partial R_i} \Bigg]\:.\numberthis \label{equ8_Reorg}
\end{align*}
The first and the second term on the right-hand side represent the force due to the change in the projectors as a result of the ionic 
displacements, while the third term represents that due to a change in the mutual overlap of the projectors. If the projectors are localized 
orbitals centred on the atoms, then the third term is exclusively due to the relative motion of the atoms that define the subspace. 
The first term may be written as ${\rm Tr}[\hat{\rho}\hat{X}]$, defining the operator 
$\hat{X}=\ket{\varphi_{n}}O^{nm}\bra{\partial\varphi_m / \partial R_i}$. Similarly, the second term on the right-hand side in 
Eq.~(\ref{equ8_Reorg}) is $\rm{Tr}[\hat{X^\dagger}\hat{\rho}]$. For the calculation of this latter see reference 
[\onlinecite{PhysRevB.85.085107}].

In order to evaluate the third term we shall use the following matrix identity for invertible matrices $\mathbf{M}$,
\begin{align*}
\mathbb{0} =
  \frac{d}{dR_i} \left[ \mathbb{1} \right] &{}= 
  \frac{d}{dR_i}\left[\mathbf{M}\mathbf{M}^{-1}\right]
=\frac{d\mathbf{M}}{dR_i}\mathbf{M}^{-1}+\mathbf{M}\frac{d\mathbf{M}^{-1}}{dR_i}\\
\Rightarrow  \frac{d\mathbf{M}^{-1}}{dR_i}
&{}=
- \mathbf{M}^{-1}\frac{d\mathbf{M}}{dR_i}\mathbf{M}^{-1}\:,\\
\end{align*}
where $\mathbb{0}$ is the null matrix. By applying this identity to the overlap matrix $\mathbf{O}$, the third term of Eq.~(\ref{equ8_Reorg}) 
can be rewritten as
\begin{align*}
  &{}\braket{\varphi_m|\hat{\rho}|\varphi_{n}}\frac{\partial O^{nm}}{\partial R_i}\\
  =-&{}\braket{\varphi_m|\hat{\rho}|\varphi_{n}}O^{n n'}\left[\braket{\frac{\partial\varphi_{n'}}{\partial R_i}|\varphi_{m'}}+\braket{\varphi_{n'}|\frac{\partial\varphi_{m'}}{\partial R_i}}\right]O^{m' m}\\
  =-&{}\braket{\varphi_{n'}|\frac{\partial\varphi_{m'}}{\partial R_i}}O^{m' m} \braket{\varphi_m|\hat{\rho}|\varphi_{n}}\braket{\varphi^{n}|\varphi^{n'}}+ 
  \textit{c.c.}\\
  =-&{}{\rm Tr}[\hat{X}^\dagger\hat{\rho}\hat{\mathbb{P}}]+   \textit{c.c.}\numberthis\:, \label{equ10}
\end{align*}
where the projectors obey $\hat{\mathbb{P}}^\dagger \hat{\mathbb{P}}= \hat{\mathbb{P}} \hat{\mathbb{P}}= \hat{\mathbb{P}}$.

If we now bring all the terms together, the nonorthogonality-respecting Pulay force will be given by the remarkably simple expression
\begin{align*}
F_{i}^\textit{c}
  =&{}-V_\textit{c} \mathrm{Tr}\big[\hat{\rho}\hat{X} +\hat{X}^{\dagger} \hat{\rho} - \hat{X}^\dagger\hat{\rho}\hat{\mathbb{P}}-\hat{\mathbb{P}} \hat{\rho}  \hat{X}\big] \\
  =&{}-2V_\textit{c} \Re \mathrm{Tr}\left[ \hat{\rho} \hat{X}
  \left(\hat{\mathbb{1}}-\hat{\mathbb{P}}\right)
 \right]\:.\numberthis \label{equ11}
\end{align*}
The final factor, $(\hat{\mathbb{1}}-\hat{\mathbb{P}})$, in this expression is a projector onto the space \rededit{\emph{complementary}} to the 
constrained one. The effect of variable nonorthogonality thus becomes clear. It generates an extra projection factor that cancels any 
component of the Pulay force associated with orbital derivatives that are not related to changes in the projected subspace. In other words it
cancels contributions related to changes that cannot cause a variation of the measured occupancy. 
If the operator $\hat{X}$ applies a linear transformation among the projector orbitals, then $\hat{X} = \hat{X} \hat{\mathbb{P}}$ and the 
Pulay force will vanish entirely. In contrast, if $\langle \partial \varphi_m / \partial R_i \rvert \varphi_n \rangle = 0$ for all $m$ and $n$, 
then $ \hat{X} \hat{\mathbb{P}} =  \hat{\mathbb{0}} $ and the expression will reduce to the ordinary Pulay force. 
It is possible that the projection factor in Eq.~(\ref{equ11}) is a useful addition to Pulay force calculations \emph{in general}, since even 
when orbitals nonorthogonality is not expected to arise or vary, numerical noise may cause slight variations from the condition 
$\langle \partial \varphi_m / \partial R_i \rvert \varphi_n \rangle = 0$. An example where this may arise is in force calculations
involving atom-centred pseudopotentials defined on a radial grid, which are projected onto a real or reciprocal-space Cartesian grid 
prior to integration with Kohn-Sham states.

\subsection{Implementation and procedure for  calculation}\label{ProcedureForCalculation_RE}

We have implemented the nonorthogonality Pulay forces in the linear-scaling DFT code {\sc onetep}~\cite{CKSPDHAAMMCP}, 
which uses  strictly localized, variationally-optimized nonorthogonal generalized Wannier functions (NGWFs)~\cite{PhysRevB.66.035119,MOSTOFI2002788,PhysRevB.85.193101}, ${\ket{\phi_\mu}} $, as a basis set. The NGWFs are, in turn, expressed as a 
linear combination of highly localized orthonormal psinc functions, which are essentially Fourier transforms of plane waves specified 
with a maximum cutoff energy. For a given DFT calculation, {\sc onetep} optimizes the NGWFs using a conjugate-gradients (CG) 
method in order to minimize the total energy. Within each iteration of such optimization, it minimizes~\cite{0953-8984-20-29-294207} 
the total-energy functional with respect to the density kernel $K^{\alpha\beta}, $ which builds the single-particle density matrix by means 
of $\rho(\mathbf{r},\mathbf{r'})=\phi_\alpha(\mathbf{r})K^{\alpha\beta}\phi_\beta(\mathbf{r'})$\cite{0953-8984-20-29-294207}. Thus, for a 
geometry optimization in presence of a constraint of the form contained in Eq.~(\ref{equ2_ReorgEn}), we run the following nested optimization 
loops,
\begin{enumerate}
\item Optimization of the ionic geometry,
\item Conjugate gradients optimization of the NGWFs ${\ket{\phi_\alpha}} $ within
ensemble DFT,
\item Conjugate gradients optimization of the Lagrange multiplier, $V_\textit{c}$, 
\item  Optimization of the density kernel $K^{\alpha\beta}$ within ensemble DFT.
\end{enumerate}

\rededit{We note that, although we use the NGWFs as optimized basis functions and as cDFT projectors in this work, the expression for the Pulay forces remain valid for any nonorthogonal set of projector functions.} The scheme that we follow for calculating the reorganization energy of a pentacene molecule adsorbed on a flake of graphene 
can be summarized as follows:
\begin{enumerate}
\item Optimize the geometry of the neutral system and calculate the GS energy with a DFT run. This gives the geometry 
$G0$ and the energy $E^0_{@G0}$.
\item Run cDFT for singly ionized pentacene at the geometry $G0$ in order to obtain the energy $E^+_{@G0}$.
\item Run a constrained geometry optimization to find the nuclear coordinates for the charged pentacene and the corresponding energy. 
This gives us a geometry $G+$ and an energy $E^+_{@G+}$.
\item Run DFT on neutral pentacene with geometry $G+$ to find the energy $E^0_{@G+}$ of the neutral configuration at the geometry 
of the charged state.
\end{enumerate}

The reorganization energy $\lambda$ is then given by
\begin{align*}
\lambda &{}=\lambda^0+\lambda^+,\\
&{}=(E^0_{@G+}-E^0_{@G0})+(E^+_{@G0}-E^+_{@G+}).\numberthis \label{eqnReorgEqn}
\end{align*}

\rededit{Geometry relaxation is performed only on the pentacene molecule, keeping the graphene flake fixed. In other words, the reorganization energies so obtained correspond to pentacene only.} Our calculations have been performed with the Perdew-Burke-Ernzerhof (PBE)~\cite{PhysRevLett.77.3865} parameterization of the 
generalized-gradient approximation of exchange-correlation functional and norm-conserving pseudopotentials. The NGWF cutoff radius 
was set to 9~$a_0$. It was found that a very high plane-wave cutoff energy of $1500$~eV is needed to avoid small changes in energy due 
to the egg-box effect. cDFT optimization is performed with conjugate gradient with the convergence threshold of $10^{-5}$~e/eV for the 
Lagrange multiplier gradient. This translates to an error of $<4 \times 10^{-4}\%$ in the population of the pentacene molecule. Geometry 
relaxation is performed with a quasi-Newton method~\cite{PFROMMER1997233} using a Broyden-Fletcher-Goldfarb-Shanno 
algorithm~\cite{PhysRevB.83.195102} with Pulay corrected forces (including correction for any 
residual NGWF non-convergence~\cite{doi:10.1063/1.4728026}) and an energy convergence threshold of $2.5 \times10^{-6}$ eV per 
atom. Some additional features employed in our calculations are described in the following subsections.

A numerical evaluation of the orbitals used to construct the constrained pentacene subspace follows below, but as standard we have 
adopted the well-established practice~\cite{PhysRevB.93.165102,PhysRevB.82.081102,Lukman,DNDmMKVIA,0953-8984-24-41-415603,PhysRevB.74.125120} of using  Wannier 
functions centred on the appropriate atoms (in this case the pentacene carbon and hydrogen) for the projectors $\lvert \varphi_m \rangle$. 
In particular, these were chosen as a subset of the NGWFs variationally-optimised for the valence manifold of the unconstrained, relaxed 
neutral ground-state of the pentacene-graphene system, following the protocol  proposed in  Ref.~[\onlinecite{PhysRevB.82.081102}] and 
first applied to cDFT in Ref.~[\onlinecite{PhysRevB.93.165102}].

\subsubsection*{Ensemble density-functional theory}
The occupation number of states in the vicinity of the Fermi level is ill-conditioned in the case of a high degree of degeneracy, as
in metals and near-metals like graphene. In other words, significant fluctuations in the occupation numbers and in the electron 
density take place despite tiny energy changes. In these situations the number of self-consistent steps necessary for converging 
the ground state can be large. In order to circumvent this problem, we employ the finite temperature ensemble DFT (and cDFT) 
formalism~\cite{MVP} as implemented within \textsc{onetep}~\cite{ARSCKS}. Here, instead of the energy, one minimizes the 
Helmholtz free energy 
\begin{align*}
  &{}A[T,\{\epsilon_i\},\{\ket{\psi_i}\}]=\sum_{i}f_i\langle \psi_i|-\frac{1}{2}\nabla^{2}|\psi_i\rangle \numberthis \\
  &{}+\int d\mathbf{r} \; v_n(\mathbf{r})\rho(\mathbf{r})+J[\rho]+E_{xc}[\rho^{\alpha},\rho^{\beta}]-TS[\{f_i(\epsilon_i)\}]\;, \label{equFreeEn}
\end{align*}  
where $S[\{f_i\}]$ is the entropy of the system given by~\cite{mandl1971statistical}, 
\begin{equation}\label{ENTROPY}
S[\{f_i\}]=-k_B\sum_i \left[f_i {\rm ln} f_i+ \left(1-f_i \right) {\rm ln} \left(1-f_i \right)\right]\;. 
\end{equation}
Here, the occupation number $f_i(\epsilon_i)$ is that of the $i$-th KS state and it follows the Fermi-Dirac distribution 
\begin{equation}\label{FD}
f_i(\epsilon_i)=\left(1-\exp\left[\frac{\epsilon_i-\mu}{k_\mathrm{B}T}\right]\right)^{-1}\:,
\end{equation}
with $\mu$ being the chemical potential, $k_\mathrm{B}$ the Boltzmann constant and $T$ the temperature. In all our calculations
we have used $T=$~300~K.

\subsubsection*{Correction for periodic boundary conditions}
Since {\sc onetep} uses the fast Fourier transform to solve the Poisson equation, it requires using periodic boundary conditions. For 
isolated systems one then constructs artificial periodic replica  of the simulation cell. This gives rise to undesired interactions 
between the cells. In order to correct such shortfall, we have used  the Martyna-Tuckerman scheme~\cite{MT} of replacing 
the Coulomb interaction from the periodic images of the simulation cell with a minimum image convention technique. This essentially 
adds an screening potential term to approximately cancel the Coulomb interactions from neighbouring cells~\cite{HDHS}. We used the 
Martyna-Tuckerman parameter of 7.0~$a_0$ that is recommended in reference~[\onlinecite{MT}].

\subsubsection*{Dispersion correction}
Dispersion interactions, which are poorly accounted for in semi-local exchange and correlation functionals are expected to be dominant 
between the pentacene molecule and the graphene flake. Hence, we use an empirical correction, $E_\mathrm{disp}(r_{ij})$, to the total 
energy, in the form of a damped London term summing over all pairs of atoms $(i,j)$ with interatomic distance of $r_{ij}$, given by
\begin{equation}\label{DISPERSION}
E_\textrm{disp}(r_{ij})=-\sum_{i>j}f_\textrm{damp}(r_{ij})\frac{C_{6,ij}}{r_{ij}^6}\:,
\end{equation}
where $C_{6,ij}$ depends on the particular pair of atoms and the damping term is given by~\cite{MEPHTFSSEK} 
\begin{equation}\label{ELSTNER}
f_\textrm{damp}(r_{ij})=(1-\exp(-c_\textrm{damp}(r_{ij}/R_{0,ij})^7))^4\:.
\end{equation}
The parameters, $c_\textrm{damp}$ and $R^{0}_{ij}$, used here have been generated and implemented previously in the \textsc{onetep} 
code by fitting a set of $60$ complexes with significant dispersion~\cite{QHCKS}.

\section{Results}

\subsection{Test of the forces on isolated pentacene}

In order to demonstrate the role and necessity of using nonorthogonality Pulay corrections, we first present some tests on a very 
simple system consisting of one isolated, charge-neutral pentacene molecule. We run three independent geometry relaxations, 
namely
\begin{enumerate}
\item An unconstrained DFT geometry optimization starting from an idealized initial guess for the ionic geometry of the neutral molecule. 
This provides a benchmark level of geometry optimization performance on the test system.
\item A geometry optimization with the same initial guess of  1., while applying a fixed constraint potential of strength $V_\textit{c}$ to 
the pre-defined pentacene space and relaxing without the force correction for the derivative of projector overlap [i.e. by omitting 
the last term on the right-hand side of Eq.~(\ref{equ8_Reorg})].
\item The same relaxation of 2., but including the exact expression of the Pulay forces given in Eq.~(\ref{equ8_Reorg}).
\end{enumerate}

\begin{figure}
\centering
\includegraphics[width=\columnwidth]{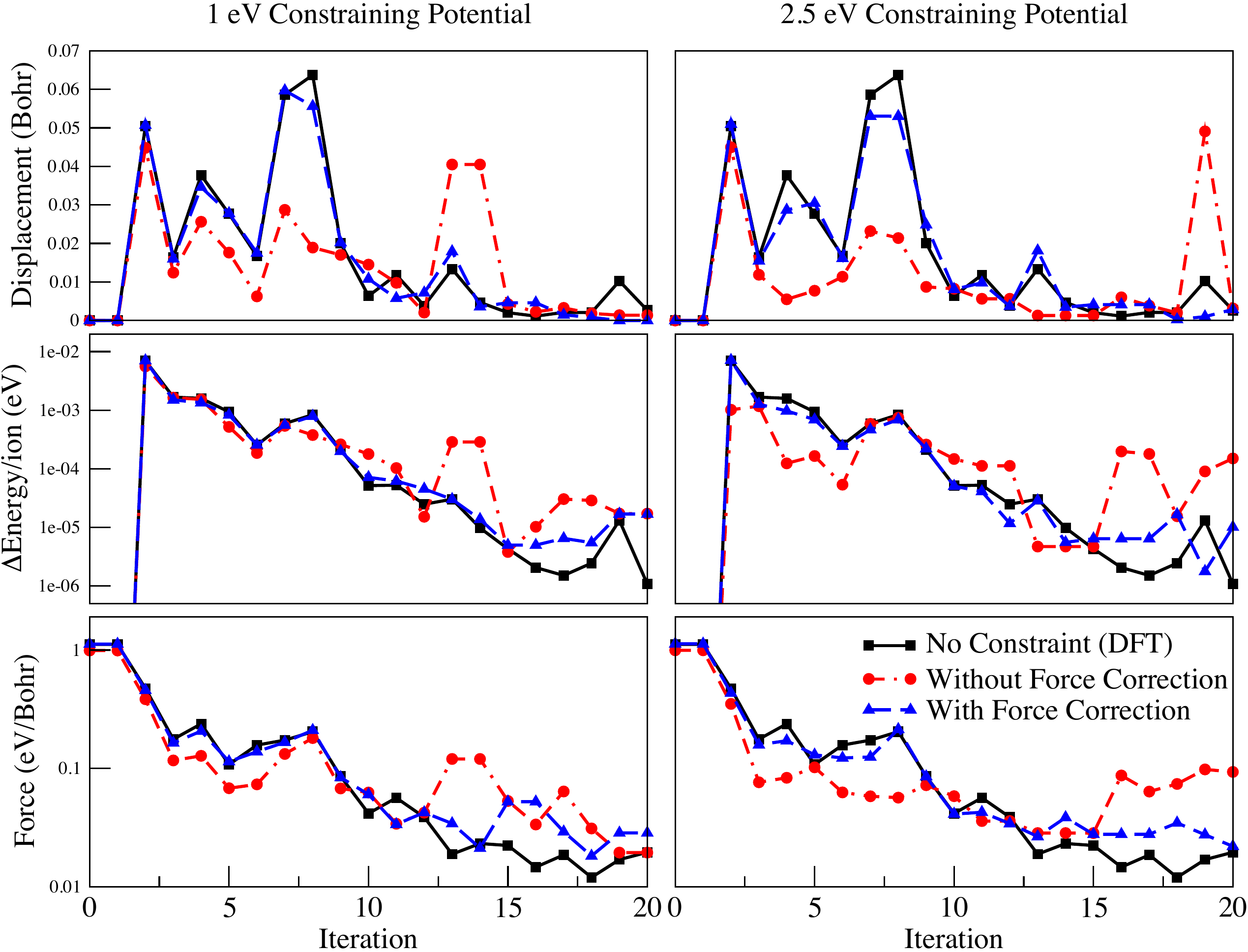}
\caption {The maximum displacement of any atom, the change in energy per atom, and the maximum force on any atom, 
plotted against the iteration number in a geometry relaxation calculation. The black, blue and red curves show the plots for 
a regular DFT run, a constrained run without the properly corrected forces and one with the proper correction for forces, 
respectively. The constrained calculations are separately performed with a fixed constant potential $V_\textit{c}=1$~eV 
(left column) and $V_\textit{c}=2.5$~eV (right column). See main text for details.}\label{FIGForceTest}
\end{figure}

A fixed, minimal set of valence pseudo-orbitals (the initial guesses for the NGWFs prior to optimization, i.e. 
H 1$s$ and C 2$p$ and 2$p$) were used to define the constrained subspace, with tensorially consistent population analysis.
In Fig.~\ref{FIGForceTest} we plot the maximum displacement, the change in energy per ion and the maximum force as a function 
of the iteration number for the aforementioned calculations performed with two different $V_\textit{c}$, namely $1$~eV and $2.5$~eV. 
For the $V_\textit{c}=$~1~eV the three calculations differ only slightly since the force correction is small. However, for $V_\textit{c}=2.5$~eV we 
see that the behaviour of the calculation using the incorrect force (red line) differs significantly from the other two, especially for the 
maximum force on any atom. In order to quantify the difference in force between the cDFT runs with and without force correction, we 
calculate the root-mean-squared (RMS) difference between the two quantities, given by 
\begin{equation}
\sqrt{\frac{1}{n}\sum_i\left(\frac{F_C^i-F_U^i}{F_C^i}\right)^2},
\end{equation}
where $F_C^i$ and $F_U^i$ are, respectively, the corrected and uncorrected, total ionic forces on the $i$-th iteration. Here, $n$ is 
the total number of iterations and, clearly, the potential that generates these forces differs except upon the first iteration. The atom
with the largest force may also change from iteration to iteration. In percentage terms, the RMS force differences are a very significant 
$121.73$~\% and $112.65$~\% for $1$~eV and $2.5$~eV constraint potentials, respectively.

\subsection{Reorganization energy of graphene-adsorbed pentacene}

\begin{figure}
\centering
\includegraphics[width=\columnwidth]{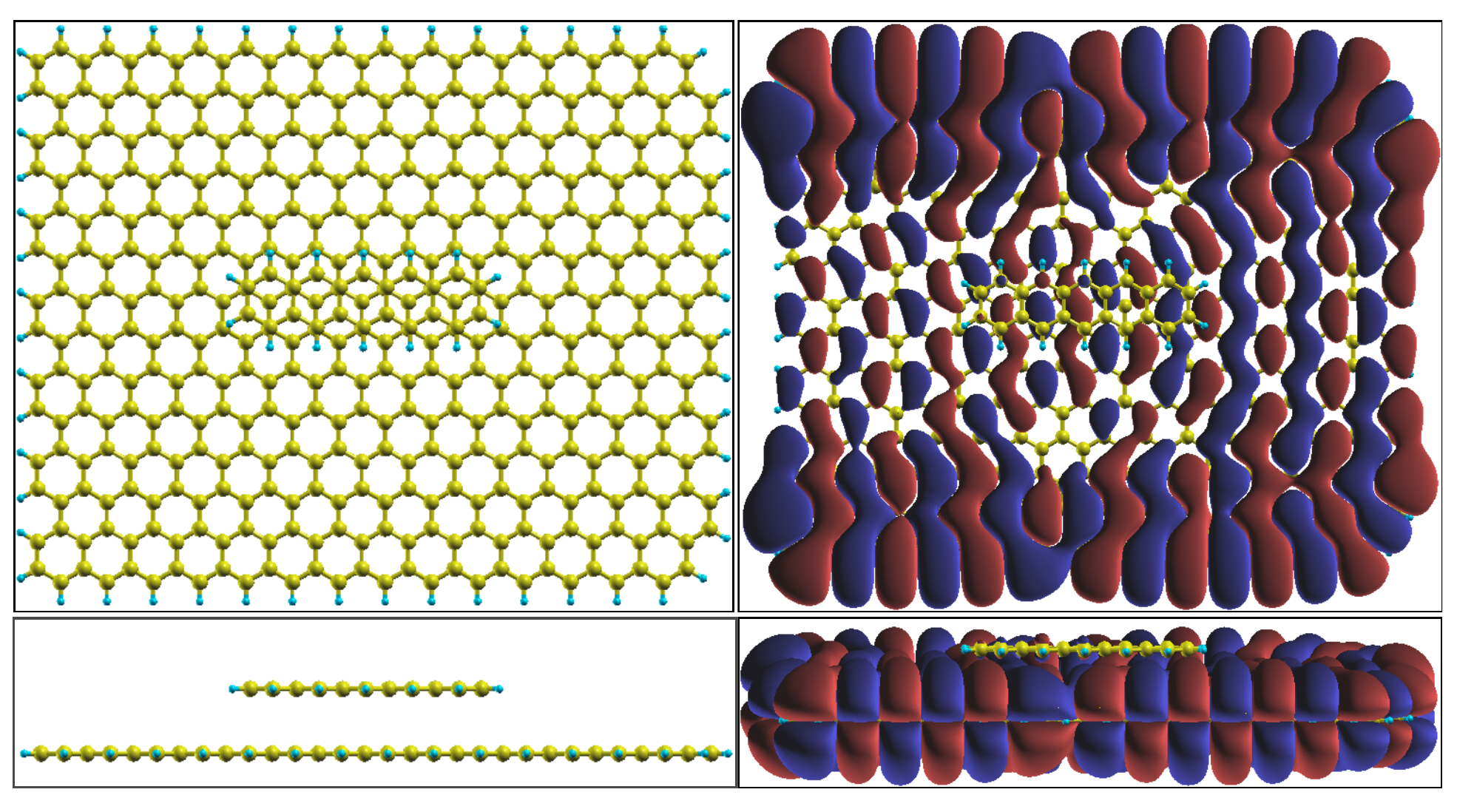}
\caption{The left-hand side panel shows the system of interest, namely a pentacene molecule adsorbed on a graphene flake. 
The right-hand side one shows an isovalue plot of the HOMO of the neutral Kohn-Sham system. It is clear that the HOMO
is confined to the graphene flake, with most of its amplitude located at its edges.}\label{FIG:SYSTEM}
\end{figure}

In this section we present and discuss our results concerning the reorganization energy of pentacene molecules adsorbed on a 
flake of graphene. The molecule is positioned above the graphene flake at its centre and is oriented parallel to it. We have performed 
our calculations with two different shapes and sizes of hydrogen-passivated graphene flake, one containing 358 atoms (hereafter referred 
to as the smaller flake) and another 474 (hereafter referred to as the larger flake). The geometry of the smaller flake has been relaxed in 
isolation. However, for the larger flake we use the geometry of an infinite graphene sheet so that the positions of the carbon atoms are 
symmetric with respect to each other, in order to better emulate an infinite graphene sheet. The system is shown in the left-hand side 
panel of Fig.~\ref{FIG:SYSTEM}, while the right-hand side one shows a plot of the highest occupied molecular orbital (HOMO) of the 
entire system. Since the HOMO is mostly localized on the graphene flake (at its edges), simply running a DFT calculation with one less 
electron is not an option as an electron will be removed from the graphene flake. Thus we use cDFT to constrain a unit positive charge on the pentacene. \rededit{We emphasize that we do not treat the reorganization effect due to pentacene-graphene charge transfer, but rather
the photoemission reorganization effect, in which 
an electron is removed from the 
pentacene molecule. This leaves the
simulation cell with a net positive charge. The monopole
interactions between the periodic replica of the charged
unit cells are neutralized by the periodic boundary correction mentioned in section~\ref{ProcedureForCalculation_RE}.}

\subsubsection{Orbital-based population analysis}\label{SubSubsection_PopAn_RE}
\begin{figure}
\centering
\subfloat[]{
  \includegraphics[width=\columnwidth]{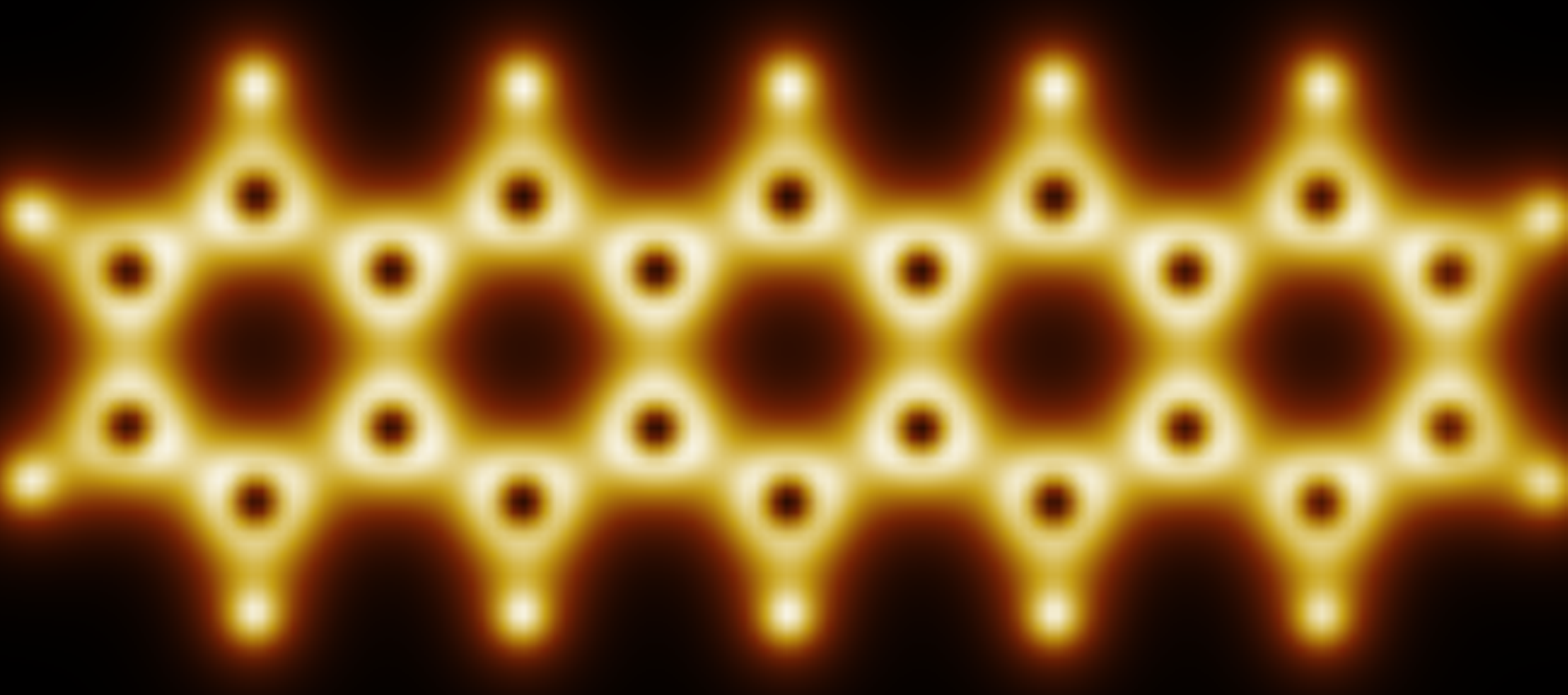}}%
  
\subfloat[]{
 \includegraphics[width=\columnwidth]{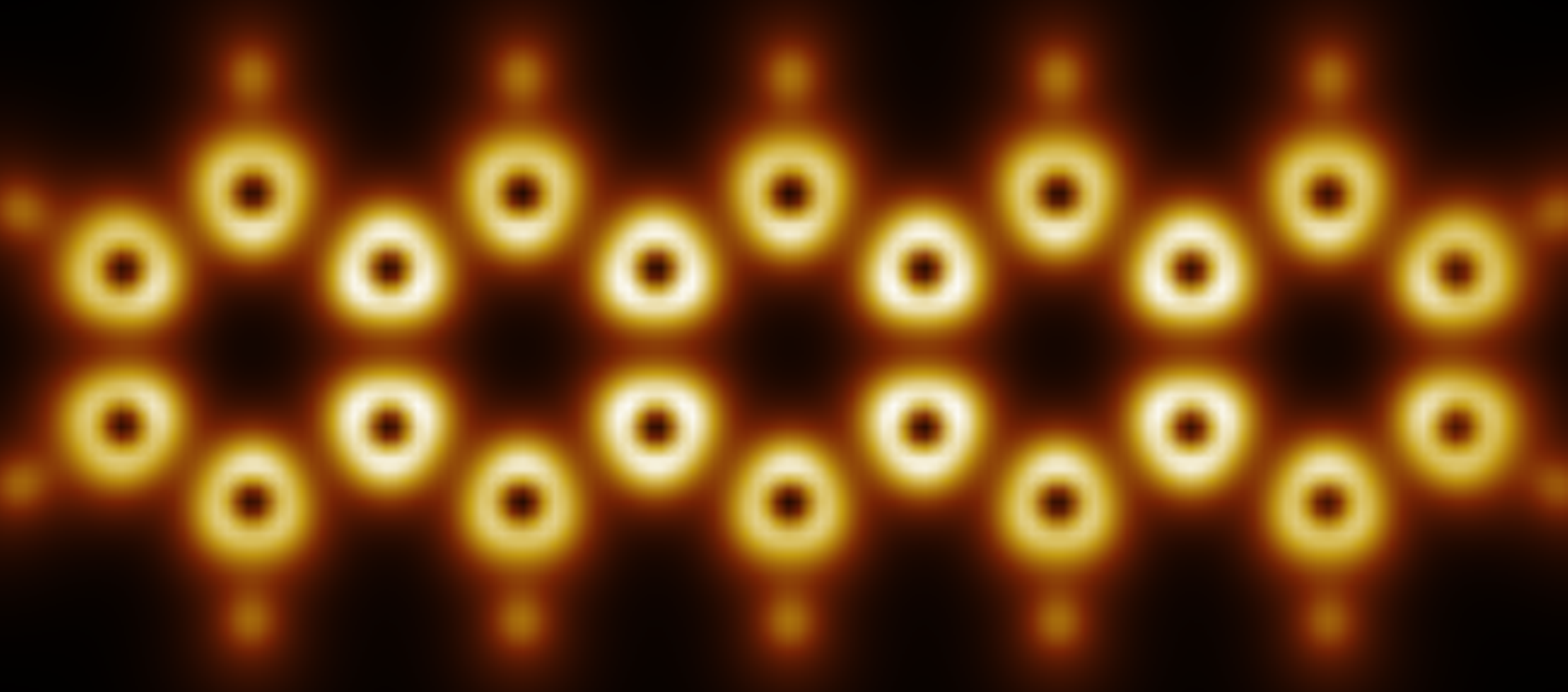}}%
\caption{Plot of $\braket{r|\hat{\mathbb{P}}|r}$ for the pentacene molecule adsorbed on graphene in the neutral state. The top and the 
bottom panels correspond respectively to the `Summed' analysis, which calculates population on individual atoms separately before adding 
them up, and the `Unified' analysis, which calculates population of the entire subspace as a whole, respectively. In the case of the Summed 
method, significant brightness in the interstitial space between atoms indicates double-counting in the region of orbital overlap. Clearly, this 
is not the case for the Unified method.}\label{fig:rPr}
\end{figure}

In the cDFT calculations we intend to remove one electron from the pentacene molecule. It is therefore necessary to carry out 
a population analysis for the uncharged ground state in order to find the number of electrons in the molecule and to define the 
constraining potential. This population depends on the choice of projectors used to represent the subspace assigned to the molecule. 
In {\sc onetep} it is possible to use as projectors the atomic pseudo-orbitals (generated from a self-consistent pseudo-atomic solver) or 
the optimized NGWFs from a previous successful run (in our case a DFT run for the same system). In both cases only the NGWFs 
associated with the relevant atoms, which here are all the pentacene atoms, are considered. Once the choice of projectors is made, 
{\sc onetep} allows predominantly two kinds of population analysis on the set of target atoms. The first technique (the `Summed' analysis) 
essentially calculates the populations on each individual atom and then sums them up. This population is defined as 
\begin{align}\label{Summed_PopAn}
N_\textrm{Summed}=\sum\limits_{I}\sum\limits_{m m'}\bra{\varphi^I_{m}}\hat{\rho} \ket{\varphi^I_{m'}}O_I^{m'm}\:,
\end{align}
where $I$ is an atom in the desired set and $O_I^{m'm}$ are the elements of the inverse of the overlap matrix of the projectors 
$\ket{\varphi^I_{m}}$ and $\ket{\varphi^I_{m'}}$ belonging to atom $I$ (this is very close to a Kronecker delta matrix in the case 
of the pseudo-atomic orbitals). The second one (the `Unified' technique) calculates the tensorially invariant population 
of the entire subspace  as 
\begin{align}\label{Unified_PoPAn}
N_\textrm{Unified}=\sum\limits_{m m'}\bra{\varphi_m}\hat{\rho}\ket{\varphi_{m'}}O^{m'm}\:,
\end{align}
where the sum is over all the orbitals of the given subspace and the inverse overlap matrix is constructed accordingly~\cite{PhysRevB.93.165102,ORPM}. 
The `Unified' technique is expected to be much more reliable, since the other double-counts the population shared by the projectors 
belonging to different atoms. This is clearly seen in Fig.~\ref{fig:rPr}, which shows a plot of $\braket{r|\hat{\mathbb{P}}|r}$ for the neutral 
pentacene molecule adsorbed on the graphene flake, where the positions $r$ lie on a plane passing close to all of the pentacene atoms. 
Using the Summed scheme (top panel) we see significant positive values of $\braket{r|\hat{\mathbb{P}}|r}$ in the interstitial region between 
the atoms, indicating the aforementioned double-counting. As expected, this is not present in the plot for the Unified scheme (bottom panel). 

   \begin{table}
   \centering
    \begin{tabular}{ ccc}
    \hline \hline \vspace{0.5mm}
    Projector & Analysis & Population \\ \hline 
    Atomic orbitals & Summed & 171.56\\ 
    Atomic orbitals & Unified & 100.74\\ 
    Optimized NGWFs & Summed & 172.72\\ 
    Optimized NGWFs & Unified & 102.11\\ \hline \hline
    \end{tabular}
     \captionsetup{justification=justified, singlelinecheck=false}
     \caption{Number of electrons on the pentacene molecule calculated by using different choices of projectors and for different population 
     analysis methods. An isolated pentacene molecule has 102 valence electrons. Please see text in section~\ref{SubSubsection_PopAn_RE} 
     for the definitions of `Summed' and `Unified'.}
       \label{table:PopulationAnalysis}
    \end{table}

In Tab.~\ref{table:PopulationAnalysis}, we tabulate the populations calculated with the different techniques/projectors on the pentacene 
molecule, which is adsorbed on a flake of graphene. Noting that an isolated pentacene molecule has 102 valence electrons, we see that 
the combination of optimized NGWFs with the Unified scheme reproduces this count to $0.1$\%, and  so we use this population analysis 
for further calculations. The residual  $0.1$\% is due to hybridization with the graphene substrate (a very slight chemisorption effect).
We note that pseudo-atomic population analysis exhibits an under-count of approximately $1$\%, but that this is small compared to the error 
of `Summed', or sometimes known as `on-site', population analysis. The significance of this result is that even pseudo-atomic orbitals can 
provide a reasonable population analysis device for cDFT, if the nonorthogonality or equivalent L\"owdin treatment is tensorially invariant 
(if it uses $\mathbf{O}$).

\subsubsection{Calculation of the reorganization energy}

\begin{figure}
\centering
\subfloat[]{
  \fbox{\includegraphics[width=\columnwidth]{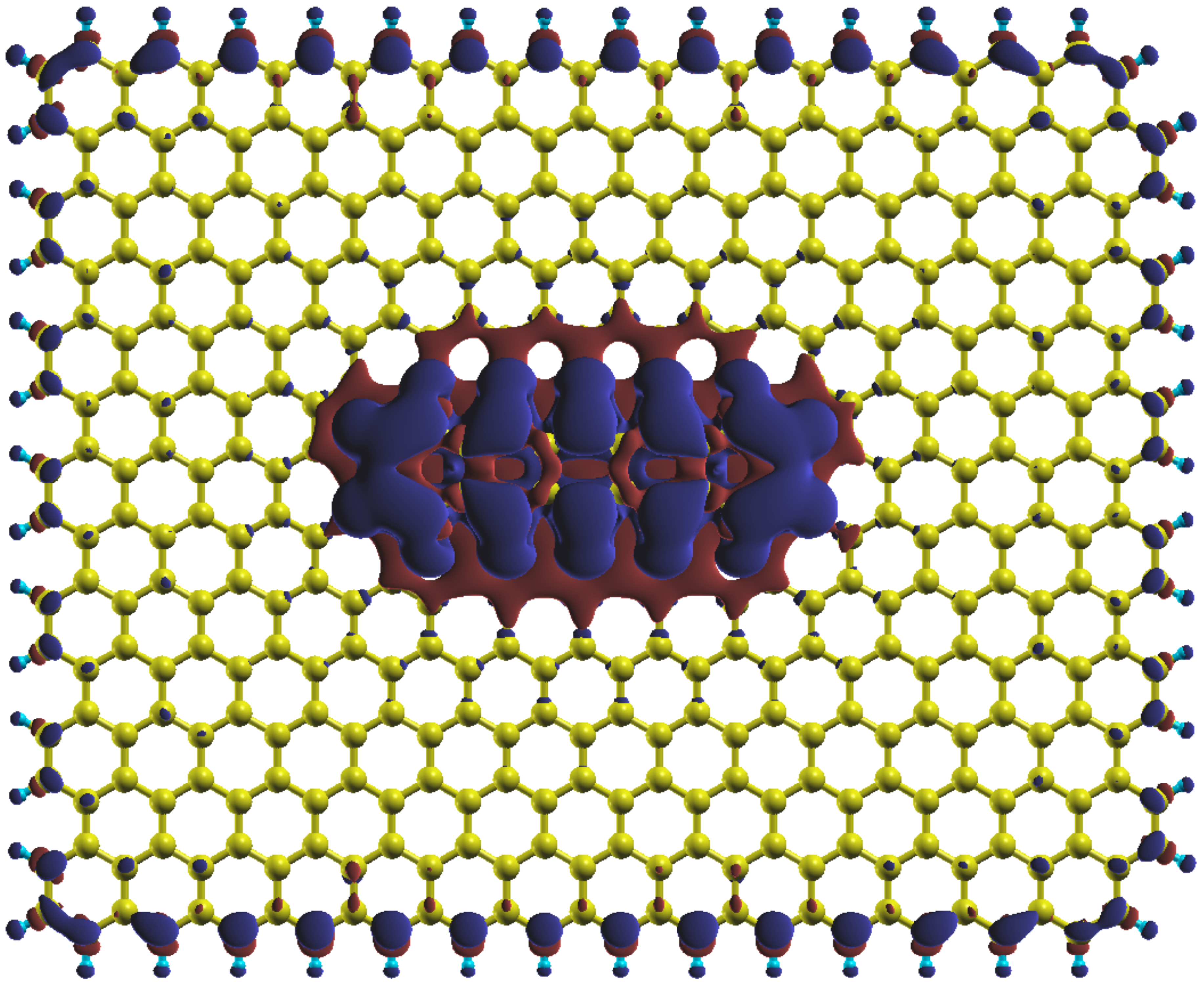}}%
}

\subfloat[]{
  \fbox{\includegraphics[width=\columnwidth]{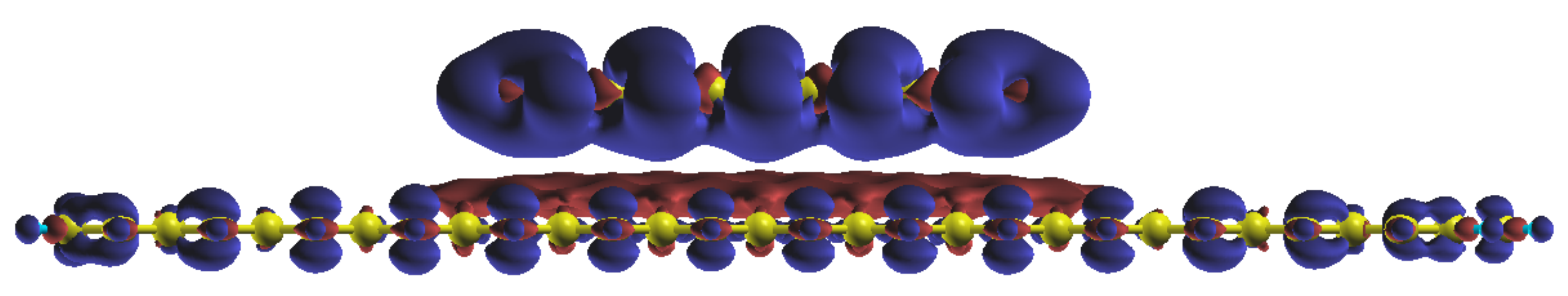}}%
}
\caption{Plot of isovalues of the change in charge density upon the removal of an electron from the molecule as calculated with cDFT. 
Blue and red colors denote positive and negative charge densities, respectively.}\label{FIG:DensityPlot}
\end{figure}

Once the population of the molecule, $N$, is determined, the target population for the cDFT calculation is defined as 
$N_\textit{c} = N \times 101/102$. Fig~\ref{FIG:DensityPlot} shows the charge density on the system after the removal of an 
electron from the molecule. As seen in the picture, a molecule with a net positive charge induces a negative charge in the region 
of the graphene flake immediately beneath the molecule. This is the image charge. 

We follow the steps outlined in the subsection~\ref{ProcedureForCalculation_RE} to calculate the reorganization energy of the 
pentacene molecule adsorbed on the graphene flake. \rededit{Since the final energy of a {\sc onetep} calculation is dependent, albeit very weakly, on the initial NGWFs, we ensure that both the calculations used for computing each instance of $\lambda^0$ or $\lambda^+$ use optimized NGWFs of as similar a provenance as possible.} The main problem with such calculation is the existence of multiple configurational 
local minima differing only slightly in energy. The local minimum that a structural relaxation converges to depends largely on the 
initial geometry. Therefore we find the reorganization energy corresponding to the two local minima (one for the uncharged system and 
another for the charged one).

\begin{figure}
\subfloat[]{
  \fbox{\includegraphics[width=\columnwidth]{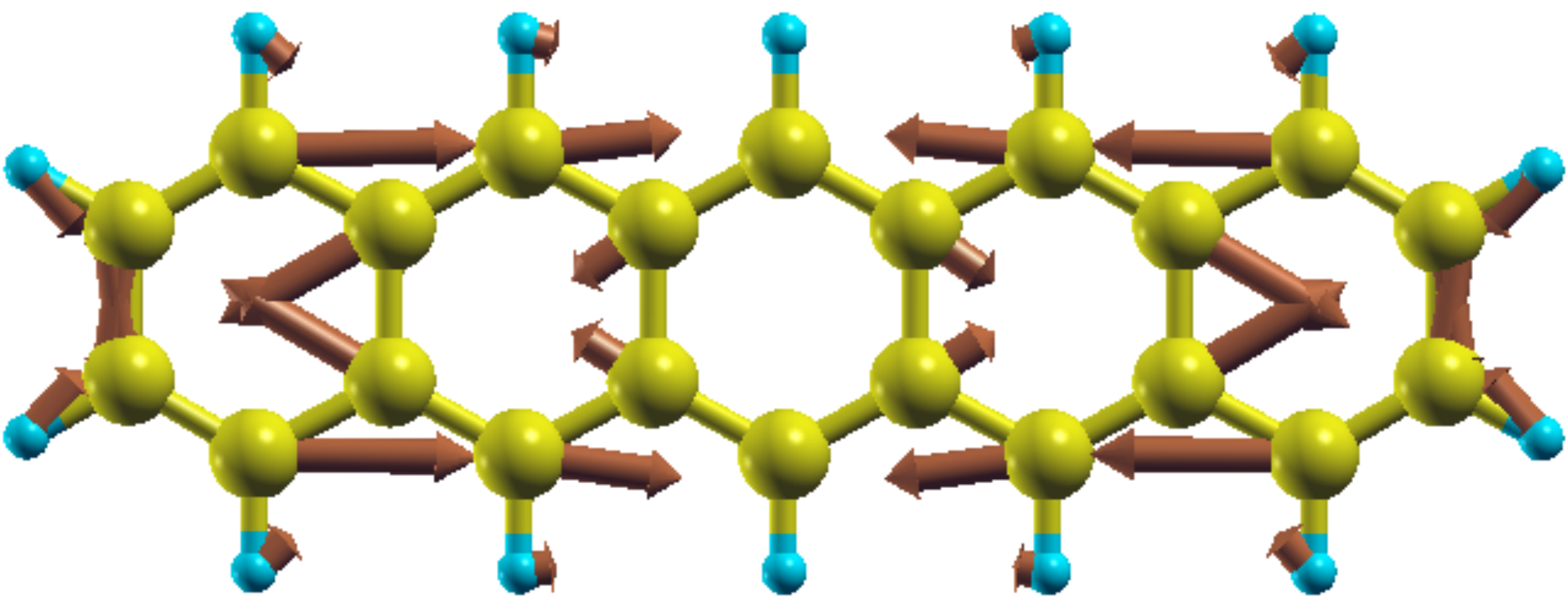}}%
}
\vspace{-0.620cm}
\subfloat[]{
  \fbox{\includegraphics[width=\columnwidth]{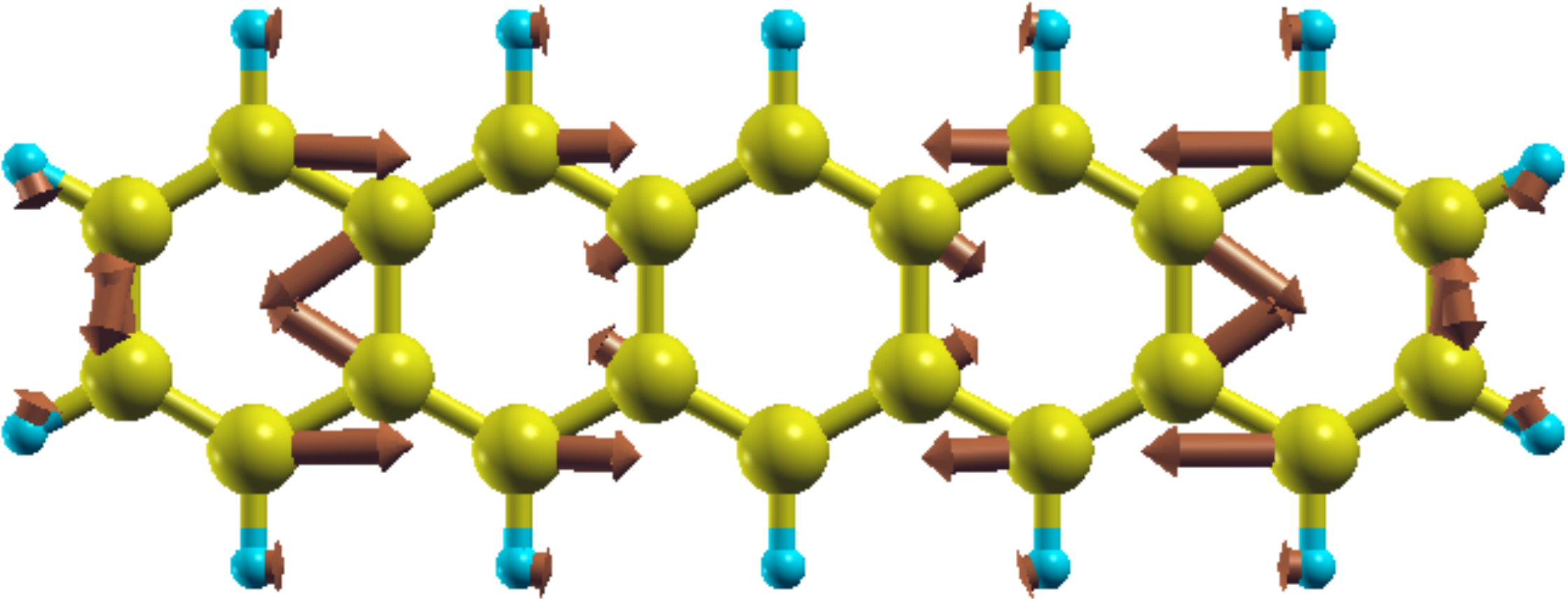}}%
}
\caption{Pentacene molecular geometry. The arrows show the directions and magnitudes of the in-plane displacement of the atoms in 
response to the removal of one electron. The top and the bottom figures correspond to an isolated pentacene and one deposited on a 
graphene flake, respectively. The graphene substrate introduces an effective steric hindrance, which reduces the reorganization effect
and energy}\label{fig:DISPLACEMENTS}
\end{figure}

As the opposite image charge formed on the flake results in a Coulomb attraction between the molecule and the flake, in the charged 
state geometry $G_+$, the molecule is closer to the flake than in the uncharged geometry $G_0$. We also 
notice that the directions of the in-plane displacements of the atoms of the pentacene molecule upon charging are very similar for the 
isolated molecule and for the molecule adsorbed on the graphene flake, as it can be seen in Fig.~\ref{fig:DISPLACEMENTS}. Furthermore, 
the average bond-length of the relaxed pentacene molecule is smaller for the charged case, for both the isolated and the adsorbed molecule. 
This indicates a shrinking of the molecule on electron emission. Such change in the average bond-length is larger for the isolated pentacene 
than for the adsorbed one, as indicated by length of the arrows in Fig.~\ref{fig:DISPLACEMENTS}. This can be attributed to the steric effects 
due to the presence of graphene.
However, as mentioned earlier, one must keep in mind that these properties can, in principle, be specific 
to the pair of local geometry minima pertaining the calculation. For a different pair of minima, these values could be different in principle. 

   \begin{table}
   \centering
    \begin{tabular}{ cccccc}
    \hline \hline \vspace{0.5mm}
    Cutoff energy & flake & \hspace{2.5mm} $\lambda^0$ \hspace{2.5mm} & \hspace{2.5mm} $\lambda^+$ \hspace{2.5mm} & \hspace{2.5mm} $\lambda$ \hspace{2.5mm} & \hspace{0.5mm} $\Delta V_\textit{c}$ 
    \hspace{1.5mm} \\ \hline 
    900 eV & none & 29 & 27 & 56 & N.A.\\
    900 eV & smaller & \rededit{23} & \rededit{26} & \rededit{49} & \rededit{44}\\ 
    900 eV & larger & \rededit{20} & \rededit{20} & \rededit{39} & \rededit{50}\\
    1500 eV & none & 29 & 27 & 56 & N.A.\\
    1500 eV & smaller & \rededit{25} & \rededit{25} & \rededit{51} & \rededit{45}\\
    1500 eV & larger & \rededit{17} & \rededit{23}  & \rededit{40} & \rededit{33}\\ \hline \hline
    \end{tabular}
     \captionsetup{justification=justified, singlelinecheck=false}
     \caption{Reorganization energies (corresponding to local minima in the geometry), of a pentacene molecule as a function of the 
     cutoff energy and the size of the graphene flake. $\Delta V_\textit{c}$ denotes the difference in the cDFT Lagrange multipliers corresponding 
     to the two different geometries. All energies are in meV. The smaller and the larger flakes contain 358 and 474 atoms
     at the optimized and idealized positions, respectively.}
       \label{table:ReorganizationEnergy}
    \end{table}

In Tab.~\ref{table:ReorganizationEnergy} we summarize our results for the reorganization energy for two different cutoff-energies and 
different sizes of the graphene flake. We have also included the reorganization energy of an isolated pentacene molecule (flake=none) 
for comparison. Note that our results for isolated pentacene matches with that obtained with MP2 method in an earlier theoretical 
study~\cite{Gruhn_SF_B_T_M_M_C_K_B}. As mentioned in Eq.~(\ref{equReOr}) here, $\lambda^0$, $\lambda^+$ and $\lambda$ refer 
to the reorganization energy contributions from the uncharged molecule, the positively charged molecule, and the total reorganization 
energy, respectively. 

\rededit{Since the reorganization energy is very small in general, minute fluctuations (per atom) arising due to diverse local geometry minimum or differences in the  NGWF initial state  can change the results considerably. As a result of exhaustive calculations using different NGWF restart protocols, we estimate the root mean square value of error caused by such deviations to be approximately $6$~meV for each instance of $\lambda^0$ or $\lambda^+$. Therefore, in Tab.~\ref{table:ReorganizationEnergy}, we  focus predominantly on the general trend in the results, which we consider to be quite robust, rather than the precise values.  
An surprising effect to observe here is that while the
total reorganization energy $\lambda$ appears to be
insensitive to changes in the kinetic cutoff
energy, relative to its separate components
$\lambda^0$ and $\lambda^+$. It is not possible to 
conclude that this is more generally the case based
on the available evidence.
The take home message of the table is} that the reorganization energy of the isolated molecule is generally greater than that of the same molecule on graphene. This can be attributed to steric effects for the latter case, namely to the fact that an adsorbed molecule has less freedom for ionic relaxation.

\rededit{The reorganization energy is lower for the larger flake. We attribute this to  two possible mechanisms: (i)  the freedom of ionic motion of the molecule may be more restricted for a larger substrate; (ii) since, as mentioned earlier, the bond lengths in the smaller flake are not all equal, adsorption on this flake is likely to result in a more uneven energy landscape for the pentacene molecule.} It is worth noting that we have analysed the different contributions due to Hartree, exchange and correlation, pseudopotentials, and kinetic energy 
to the reorganization energy. However, the relatively small rereorganization energy turns out to be the result of the substantial cancellation 
of large variations in these individual terms. It is noteworthy that experimental studies~\cite{KERA2009135,YH_NS_FH_KS_FR_OKK_UN} 
on a rather different system of graphene-adsorbed pentacene, namely a thin film of pentacene deposited on HOPG, conversely exhibits an 
increase in reorganization energy with respect to the isolated pentacene molecule. This points to the possibility that intermolecular relaxation 
in the film contributes to the reorganization energy and more than compensates for the effects of steric hindrance.

\rededit{Here we note that, since, strictly speaking, the polarizability of the neutral molecule is different from that of the charged one, using the same form of empirical vdW correction for the molecule-flake interface in both cases may introduce some bias in the numerical results. To obtain an estimate for such error, we calculate, without using any vdW correction, the reorganization energy of pentacene adsorbed on the smaller flake using plane-wave cutoff of $900$~eV. We see that the results so obtained ($\lambda^0=23$~meV and $\lambda^+=29$~meV) are similar to those obtained with vdW corrections, and that the difference is
within the range of fluctuations caused by local minima and in the NGWF restart protocol.  We infer that the inclusion of the vdW corrections does not alter the reorganization energy significantly.}

We finally note that, since the Lagrange multiplier $V_\textit{c}$ for one-electron removal may be interpreted as an unscreened approximate
subspace-local ionization potential, and since the extent of the screening may be assumed to be independent of small changes in the ionic 
geometry, so the difference, $\Delta V_\textit{c}$, between the converged Lagrange multipliers for the charged pentacene in geometries 
$G_0$ and $G_+$ can be taken as an approximation for the reorganization energy. Also, since this quantity is evaluated explicitly only
on the basis of the occupancy of the adsorbate, we may expect it to be relatively free (that is, compared to the true reorganization
energy) from numerical errors in the optimized ionic positions of distant atoms in the graphene flake. Consequently, in 
Tab.~\ref{table:ReorganizationEnergy}, we find that $\Delta V_\textit{c}$ is slightly less dependent on the nature of the substrate than the true 
reorganization energy is but, in contrast, it seems to be too sensitive to the plane-wave energy cutoff for practical utility.

\section{Conclusion}
We have presented a method for calculating self-consistent forces in conjunction with constrained DFT in first principles calculations 
employing atom-centred functions. We have investigated a very accurate population analysis constructed over Wannier functions 
and a tensorially consistent treatment of nonorthogonality. This is shown to yield an exact expression for force containing a Pulay term for 
the change in nonorthogonality, which circumvents the need for overlap matrix diagonalisation and is compatible with complex-valued orbitals. 
We have implemented this expression for the force in the DFT code {\sc onetep} and have shown that the contribution to the force arising 
from the change in mutual overlap of the nonorthogonal projector orbitals of the subspace exerts significant influence on the geometry relaxation. 

In order show a novel practical application of such forces, we perform a hyper-accurate geometry optimisation with numerous extra features to 
capture the reorganization energy of a pentacene molecule adsorbed on a flake of graphene. We have argued that the Lagrange multiplier itself 
can be used to provide a local estimate of the reorganization energy in systems, where the principal change to the system is spatially localised. 
Since the geometry of such system has multiple local minima closely related in energy, the reorganization energy can, in principle, be calculated 
only over such local minima. These depend on the initial geometry. We show that for the minima obtained in our calculations, the reorganization 
energy of the molecule adsorbed on a graphene flake is typically smaller than that of the isolated molecule, a fact that is consistent with a steric hindrance effect.

\section*{ACKNOWLEDGEMENTS}

This work is supported by the European Research Council project \textsc{Quest}. 
We acknowledge and thank G. Teobaldi and N. D. M. Hine for their implementation and automation of cDFT in \textsc{ONETEP}, 
D. Turban for extending the population analysis to encompass orbitals across multiple atomic centres, and C.-K. Skylaris and 
his team for their prior implementation of the dispersion correction, boundary condition correction, and ensemble DFT in that package.
The authors  acknowledge the DJEI/DES/SFI/HEA Irish Centre for High-End Computing (ICHEC) for the provision of computational facilities and support. We also acknowledge Trinity Research IT for the maintenance of the Boyle cluster on which further calculations were performed.


%

\end{document}